\documentstyle[prl,twocolumn,aps]{revtex}

\newcommand{\be}{\begin{equation}}
\newcommand{\ee}{\end{equation}}

\begin{document}
\twocolumn[\hsize\textwidth\columnwidth\hsize\csname @twocolumnfalse\endcsname
\draft
\title{Gravitational Energy-Momentum Density in Teleparallel Gravity}
\author{V. C. de Andrade, L. C. T. Guillen and J. G. Pereira}
\address{Instituto de F\'{\i}sica Te\'orica,
Universidade Estadual Paulista \\
Rua Pamplona 145, 01405-900 S\~ao Paulo SP, Brazil}
\maketitle

\begin{abstract}

In the context of a gauge theory for the translation group, a conserved energy-momentum
gauge current for the gravitational field is obtained. It is a true spacetime and gauge
tensor, and transforms covariantly under global Lorentz transformations. By rewriting
the gauge gravitational field equation in a purely spacetime form, it becomes the
teleparallel equivalent of Einstein's equation, and the gauge current reduces to the
M{\o}ller's canonical energy-momentum density of the gravitational field.

\end{abstract}


\vskip1pc]

The definition of an energy-momentum density for the gravitational field is one of the oldest
and most controversial problems of gravitation. As a true field, it would be natural to
expect that gravity should have its own local energy-momentum density. However, it is usually
asserted that such a density can not be locally defined because of the equivalence
principle~\cite{gravitation}. As a consequence, any attempt to identify an energy-momentum
density for the gravitational field leads to complexes that are not true tensors. The first
of such attempt was made by Einstein who proposed an expression for the energy-momentum
density of the gravitational field which was nothing but the canonical expression obtained
from Noether's theorem~\cite{trautman}. Indeed, this quantity is a pseudotensor, an object
that depends on the coordinate system. Several other attempts have been made, leading to
different expressions for the energy-momentum pseudotensor for the gravitational
field~\cite{others}.

Despite the existence of some controversial points related to the formulation of the
equivalence principle~\cite{synge}, it seems true that, in the context of general relativity,
no tensorial expression for the gravitational energy-momentum density can exist. However, as
our results show, in the gauge context, the existence of an expression for the gravitational
energy-momentum density which is a true spacetime and gauge tensor turns out to be possible.
Accordingly, the absence of such expression should be attributed to the general relativity
description of gravitation, which seems to be not the appropriate framework to deal with this
problem~\cite{maluf95}.

In spite of some skepticism~\cite{gravitation}, there has been a continuous interest in this
problem~\cite{recent}. In particular, a {\it quasilocal} approach has been proposed recently
which is highly clarifying~\cite{nester}. According to this approach, for each gravitational
energy-momentum pseudotensor, there is an associated {\it superpotential} which is a
hamiltonian boundary term. The energy-momentum defined by such a pseudotensor does not really
depend on the local value of the reference frame, but only on the value of the reference
frame on the boundary of a region --- then its {\it quasilocal} character. As the relevant
boundary conditions are physically acceptable, this approach validates the pseudotensor
approach to the gravitational energy-momentum problem. It should be mentioned that these
results were obtained in the context of the general relativity description of gravitation.

In the present work a different approach will be used to re-examine the gravitational
energy-momentum problem. Due to the fundamental character of the geometric structure
underlying gauge theories, the concept of currents, and in particular the concepts of energy
and momentum, are much more transparent when considered from the gauge point of
view~\cite{gh}. Accordingly, we are going to consider gravity as described by a gauge
theory~\cite{hehl}. Our basic interest will be concentrated on the gauge theories for the
translation group~\cite{trans}, and in particular on the so called teleparallel equivalent of
general relativity~\cite{maluf}. It is important to remark that this equivalence is true only
in the absence of spinor matter fields~\cite{hayashi}.

Let us start by reviewing the fundamentals of the teleparallel equivalent of general
relativity. We use the Greek alphabet $(\mu, \nu, \rho, \dots = 0,1,2,3)$ to denote indices
related to spacetime, and the Latin alphabet $(a,b,c, \dots = 0,1,2,3)$ to denote indices
related to the tangent space (fiber), assumed to be a Minkowski space with the metric
$\eta_{ab}=\mbox{diag}(+1,-1,-1,-1)$. A gauge transformation is defined as a local
translation of the tangent-space coordinates,
\begin{equation}
\delta x^{a} = \delta\alpha^{b}P_{b}x^{a},
\end{equation}
with $P_{a} = \partial /\partial x^a$ the translation generators, and $\delta \alpha^{a}$
the corresponding infinitesimal parameters. Denoting the gauge potentials by $A^{a}{}_{\mu}$,
the gauge covariant derivative of a general matter field $\Psi$ is~\cite{paper1}
\be
{\mathcal D}_\mu \Psi = h^{a}{}_{\mu} \; \partial_{a} \Psi \; ,
\ee
where
\be
h^{a}{}_{\mu} = \partial_{\mu}x^{a} + c^{-2}A^{a}{}_{\mu}  
\label{2.17}
\ee
is a nontrivial tetrad field, with $c$ the speed of light. From the covariance of 
$D_{\mu} \Psi$, we obtain the transformation of the gauge potentials:
\begin{equation}
A^{a^{\prime}}{}_{\mu} = A^{a}{}_{\mu} - 
c^{2}\partial_{\mu}\delta\alpha^{a} \; .
\end{equation}
As usual in abelian gauge theories, the field strength is given by
\begin{equation}
F^{a}{}_{\mu \nu} = \partial_{\mu} A^{a}{}_{\nu} - \partial_{\nu}A^{a}{}_{\mu} \; ,
\label{core}
\end{equation}
which satisfies the relation
\begin{equation}
[{\mathcal D}_{\mu}, {\mathcal D}_{\nu}] \Psi = c^{-2} F^{a}{}_{\mu \nu} P_a \Psi .
\end{equation}
It is important to remark that, whereas the tangent space indices are raised and lowered with
the metric $\eta_{a b}$, the spacetime indices are raised and lowered with the riemannian
metric
\be
g_{\mu \nu} = \eta_{a b} h^a{}_{\mu} \, h^b{}_{\nu} \; .
\label{gmn}
\ee

A nontrivial tetrad field induces on spacetime a te\-le\-parallel structure which is directly
related to the presence of the gravitational field. In fact, given a nontrivial tetrad
$h^{a}{}_{\mu}$, it is possible to define a Cartan connection
\be
\Gamma^{\rho}{}_{\mu \nu} = h_{a}{}^{\rho}\partial_{\nu}h^{a}{}_{\mu},
\label{carco}
\ee
which is a connection presenting torsion, but no curvature~\cite{livro}. As a natural
consequence of this definition, the Cartan covariant derivative of the tetrad field vanishes
identically:
\be
\nabla_{\nu}h^a{}_{\mu} \equiv \partial_{\nu}h^a{}_{\mu} - 
\Gamma^{\theta}{}_{\mu \nu} \, h^a{}_{\theta} = 0 .
\label{cacd}
\ee
This is the absolute parallelism condition. The torsion of the Cartan connection is
\be
T^{\rho}{}_{\mu \nu} = \Gamma^{\rho}{}_{\nu \mu} - \Gamma^{\rho}{}_{\mu \nu} \; ,
\label{2.21a}
\ee
from which we see that the gravitational field strength is nothing but torsion written in
the tetrad basis:
\be
F^{a}{}_{\mu \nu}$ = $c^{2}h^{a}{}_{\rho}T^{\rho}{}_{\mu \nu} .
\ee
The Cartan connection $\Gamma^{\rho}{}_{\mu \nu}$ and the Levi-Civita connection of the
metric (\ref{gmn}), denoted by ${\stackrel{\circ}{\Gamma}}{}^{\rho} {}_{\mu \nu}$, are
related by
\be
\Gamma^{\rho}{}_{\mu \nu} = 
{\stackrel{\circ}{\Gamma}}{}^{\rho} {}_{\mu \nu} + 
K^{\rho}{}_{\mu \nu},
\label{rela}
\ee
with
\be
K^{\rho}{}_{\mu \nu} = {\textstyle \frac{1}{2}} \left( T_{\mu}{}^{\rho}{}_{\nu} +
T_{\nu}{}^{\rho}{}_{\mu} - T^{\rho}{}_{\mu \nu} \right)
\ee
the contorsion tensor.

The gauge gravitational field Lagrangian is given by~\cite{paper1}
\be
{\cal L}_G = 
\frac{h c^{4}}{16 \pi G} \; S^{\rho \mu \nu} \; T_{\rho \mu \nu},
\label{gala}
\ee
where $h = {\rm det}(h^{a}{}_{\mu})$, and
\[
S^{\rho \mu \nu} = - S^{\rho \nu \mu} \equiv {\textstyle \frac{1}{2}} \left[ K^{\mu \nu \rho}
- g^{\rho \nu} \; T^{\theta \mu}{}_{\theta} + g^{\rho \mu} \;
T^{\theta \nu}{}_{\theta} \right] 
\]
is a tensor written in terms of the Cartan connection only. As usual in gauge theories, it is
quadratic in the field strength. By using relation (\ref{rela}), this lagrangian can be
rewritten in terms of the Levi-Civita connection. Up to a total divergence, the result is the
Hilbert-Einstein Lagrangian of general relativity
\be
{\cal L} = - \frac{c^4}{16 \pi G} \;  \sqrt{-g} \, {\stackrel{\circ}{R}} ,
\ee
where the identification $h = \sqrt{-g}$ has been made.

By performing variations in relation to the gauge field $A_a{}^\rho$, we obtain from the
gauge lagrangian ${\cal L}_G$ the teleparallel version of the gravitational field equation,
\be
\partial_\sigma(h S_a{}^{\sigma \rho}) - 
\frac{4 \pi G}{c^4} \, (h j_{a}{}^{\rho}) = 0 ,
\label{tfe1}
\ee
where $S_a{}^{\sigma \rho} \equiv h_{a}{}^{\lambda}
S_{\lambda}{}^{\sigma \rho}$. Analogously to the Yang-Mills theories~\cite{ramond}, 
\be
h j_{a}{}^{\rho} \equiv - \frac{\partial {\cal L}_G}{\partial h^a{}_{\rho}} =
\frac{c^{4}}{4 \pi G} \, h h_a{}^{\lambda} S_{\mu}{}^{\nu \rho} T^\mu{}_{\nu \lambda}
- h_a{}^{\rho} {\cal L}_G
\label{ptem1} 
\ee
stands for the gravitational gauge current, which in this case represents the energy and
momentum of the gravitational field. The term $(h S_a{}^{\sigma \rho})$ is called
{\it superpotential} in the sense that its derivative yields the gauge current
$(h j_{a}{}^{\rho})$. Because of the anti-symmetry of $S_a{}^{\sigma \rho}$ in the last
two indices, $(h j_{a}{}^{\rho})$ is conserved as a consequence of the field equation:
\be
\partial_\rho (h j_a{}^\rho) = 0 .
\label{conser1}
\ee
Making use of the identity
\be
\partial_\rho h \equiv h {\Gamma}^{\nu}{}_{\nu \rho} =
h \left( {\Gamma}^{\nu}{}_{\rho \nu} - K^{\nu}{}_{\rho \nu} \right) \; ,
\label{id1}
\ee
this conservation law can be rewritten as
\be
D_\rho \, j_a{}^\rho \equiv \partial_\rho j_a{}^\rho +
\left( \Gamma^\rho{}_{\lambda \rho} - K^\rho{}_{\lambda \rho} \right) j_a{}^\lambda = 0 \; ,
\label{conser2}
\ee
where $D_\rho$ is the teleparallel version of the covariant derivative, which is nothing but
the Levi-Civita covariant derivative of general relativity rephrased in terms of the Cartan
connection~\cite{vector}. As can be easily checked, $j_a{}^\rho$ transforms covariantly
under a general spacetime coordinate transformation, and is invariant under local (gauge)
translation of the tangent-space coordinates. This means that $j_a{}^\rho$ is a true
spacetime and gauge tensor. However, it transforms covariantly only under a {\it global}
tangent-space Lorentz transformation.  

Let us now proceed further and find out the relation between the above gauge approach and
general relativity. By using Eq.~(\ref{carco}) to express $\partial_\rho h_a{}^\lambda$,
the field equation (\ref{tfe1}) can be rewritten in a purely spacetime form,
\be
\partial_\sigma(h S_\lambda{}^{\sigma \rho}) - 
\frac{4 \pi G}{c^4} \, (h t_{\lambda}{}^{\rho}) = 0 \; ,
\label{tfe2}
\ee
where now 
\be
h t_{\lambda}{}^{\rho} =
\frac{c^{4}}{4 \pi G} \, h \, \Gamma^{\mu}{}_{\nu \lambda} \, S_{\mu}{}^{\nu \rho}
+ \delta_\lambda{}^{\rho} \, {\cal L}_G
\label{ptem2} 
\ee
stands for the teleparallel version of the canonical energy-momentum pseudotensor of the
gravitational field. Despite not explicitly apparent, as a consequence of the {\it local}
Lorentz invariance~\cite{weinberg} of the gauge Lagrangian ${\cal L}_G$, the field
equation~(\ref{tfe2}) is symmetric in $(\lambda \rho)$. Furthermore, by using
Eq.~(\ref{rela}), it can be rewritten in terms of the Levi-Civita connection only.
As expected, due to the equivalence between the corresponding Lagrangians, it is the
same as Einstein's equation:
\be
\frac{h}{2} \left[{\stackrel{\circ}{R}}_{\mu \nu} -
\frac{1}{2} \, g_{\mu \nu}
{\stackrel{\circ}{R}} \right] = 0 \; .
\ee  

The canonical energy-momentum pseudotensor $t_{\lambda}{}^{\rho}$ is not simply the gauge
current $j_a{}^\rho$ with the algebraic index ``$a$'' changed to the spacetime index
``$\lambda$''. It incorporates also an extra term coming from the derivative term
of Eq.~(\ref{tfe1}):
\be
t_\lambda{}^\rho = h^a{}_\lambda \, j_a{}^\rho +
\frac{c^{4}}{4 \pi G} \, \Gamma^{\mu}{}_{\lambda \nu} S_{\mu}{}^{\nu \rho} \; .
\label{ptem3}
\ee
We see thus clearly the origin of the connection-term which transforms the gauge current
$j_a{}^\rho$ into the energy-momentum pseudotensor $t_\lambda{}^\rho$. Through the same
mechanism, it is possible to appropriately exchange further terms between the derivative and
the current terms of the field equation (\ref{tfe2}), giving rise to different definitions
for the energy-momentum pseudotensor, each one connected to a different {\it superpotential}
$(h S_\lambda{}^{\rho \sigma})$. Like the gauge current $(h j_a{}^\rho)$, the pseudotensor
$(h t_\lambda{}^\rho)$ is conserved as a consequence of the field equation:
\be
\partial_\rho (h t_\lambda{}^\rho) = 0 \; .
\label{conser3}
\ee
However, in contrast to what occurs with $j_a{}^\rho$, due to the pseudotensor character of
$t_\lambda{}^\rho$, this conservation law can not be rewritten with a covariant derivative.

Because of its simplicity and transparency, the teleparallel approach to gravitation seems
to be much more appropriate than general relativity to deal with the energy problem of the
gravitational field. In fact, M{\o}ller already noticed a long time ago that a
satisfactory solution to the problem of the energy distribution in a gravitational field
could be obtained in the framework of a tetrad theory. In our notation, his expression for
the gravitational energy-momentum density is~\cite{moller}
\be
h t_\lambda{}^\rho = \frac{\partial {\cal L}}{\partial \partial_\rho h^a{}_\mu} \;
\partial_\lambda h^a{}_\mu + \delta_\lambda{}^\rho \, {\cal L} \; ,
\ee
which is nothing but the usual canonical energy-mo\-men\-tum density yielded by Noether's
theorem. Using for ${\cal L}$ the gauge Lagrangian (\ref{gala}), it is an easy task
to verify that M{\o}ller's expression coincides exactly with the teleparallel
energy-momentum density appearing in the field equation (\ref{tfe2}-\ref{ptem2}).
Since $j_a{}^\rho$ is a true spacetime tensor, whereas $t_\lambda{}^\rho$ is not,
we can say that the gauge current $j_a{}^\rho$ is an improved version of the
M{\o}ller's energy-momentum density
$t_\lambda{}^\rho$. Mathematically, they can be obtained from each other by
Eq.~(\ref{ptem3}). It should be remarked, however, that both of them transform
covariantly only under {\it global} tangent-space Lorentz transformations. This is,
we believe, the farthest one can go in the direction of a tensorial
definition for the energy and momentum of the gravitational field. The lack of a {\it local}
Lorentz covariance can be considered as the teleparallel manifestation of the pseudotensor
character of the gravitational energy-momentum density in general relativity. Accordingly,
we can say that, if it were possible to define a {\it local} Lorentz covariant gauge
current in the teleparallel gravity, the corresponding general relativity energy-momentum
density would be represented by a true spacetime tensor.

The results can be summarized as follows. In the context of a gauge theory for the
translation group, we have obtained an energy-momentum gauge current $j_a{}^\rho$ for the
gravitational field which transforms covariantly under spacetime general coordinate
transformations, and is invariant under local (gauge) translations of the tangent-space
coordinates. This means essentially that $j_a{}^\rho$ is a true spacetime and gauge tensor.
By rewriting the gauge field equation in a purely spacetime form, it becomes equivalent to
Einstein's equation of general relativity, and the gauge current $j_a{}^\rho$ reduces to the
canonical energy-momentum pseudotensor of the gravitational field, which coincides with
M{\o}ller's well-known expression. In the ordinary context of general relativity, therefore,
the energy-momentum density for the gravitational field will always be represented by a
pseudotensor.

According to the {\it quasilocal} approach, to any energy-momentum pseudotensor there is
an associated {\it superpotential} which is a hamiltonian boundary term~\cite{nester}.
On the other hand, the teleparallel field equations explicitly exhibit both the
{\it superpotential} and the gravitational energy-momentum complex. We see then that, in fact,
by appropriately exchanging terms between the {\it superpotential} and the current terms of
the field equation (\ref{tfe2}), it is possible to obtain different gravitational
energy-momentum pseudotensors with their associated {\it superpotentials}. In this context,
our results can be rephrased according to the following scheme. First, notice that
the left-hand side of the field equation~(\ref{tfe2}) as a whole is a true tensor, though
each one of its two terms is not. Then if we extract the spurious part from the first
term --- so that it becomes a true spacetime and gauge tensor --- and add this part to
the second term --- the energy-momentum density --- it becomes also a true spacetime and
gauge tensor. We thus arrive at the gauge-type field equation~(\ref{tfe1}), with
$(h S_a{}^{\sigma \rho})$ as the {\it superpotential}, whose corresponding expression
for the conserved energy-momentum density for the gravitational field, given by
$j_a{}^{\rho}$, though transforming covariantly only under a global tangent-space
Lorentz transformation, is a true spacetime and gauge tensor. 

The authors would like to thank FAPESP-Brazil, CAPES-Brazil and CNPq-Brazil for financial
support.

\end{document}